# 1 Investigating the Effectiveness of Variance Reduction Techniques in Manufacturing, Call Center and Cross-Docking Discrete Event Simulation Models

Adrian Adewunmi[*] and Uwe Aickelin[**]

Variance reduction techniques have been shown by others in the past to be a useful tool to reduce variance in Simulation studies. However, their application and success in the past has been mainly domain specific, with relatively little guidelines as to their general applicability, in particular for novices in this area. To facilitate their use, this study aims to investigate the robustness of individual techniques across a set of scenarios from different domains. Experimental results show that Control Variates is the only technique which achieves a reduction in variance across all domains. Furthermore, applied individually, Antithetic Variates and Control Variates perform particularly well in the Cross-docking scenarios, which was previously unknown.

## 1.1 Introduction

There are several analytic methods within the field of operational research; simulation is more recognized in contrast to others such as mathematical modeling and game theory. In simulation, an analyst creates a model of a real - life system that describes some process involving individual units such as persons or products. The constituents of such a model attempt to reproduce, with some varying degree of accuracy, the actual operations of the real workings of the process under consideration. It is likely that such a real - life system will have time - varying inputs

Adrian Adewunmi · Uwe Aickelin
Intelligent Modelling & Analysis Research Group (IMA)
School of Computer Science
The University of Nottingham
Jubilee Campus
Wollaton Road
Nottingham NG8 1BB
UK
e-mail: adrian.a.adewunmi@googlemail.com,
        uwe.aickelin@nottingham.ac.uk
[*] Corresponding author.
[**] Co – author.

and time - varying outputs which may be influenced by random events (Law 2007). For all random events it is important to represent the distribution of randomness accurately within input data of the simulation model. Since random samples from input probability distributions are used to model random events in simulation model through time, basic simulation output data are also characterized by randomness (Banks et al. 2000). Such randomness is known to affect the degree of accuracy of results derived from simulation output data analysis. Consequently, there is a need to reduce the variance associated within simulation output value, using the same or less simulation effort, in order to improve a desired precision (Lavenberg and Welch 1978).

There are various alternatives for dealing with the problem of improving the accuracy of simulation experimental results. It is possible to increase the number of replications as a solution approach, but the required number of replications to achieve a desired precision is unknown in advance (Hoad et al. 2009) , (Adewunmi et al. 2008). Another solution is to exploit the source of the inherent randomness which characterizes simulation models in order to achieve the goal of improved simulation results. This can be done through the use of variance reduction techniques.

> "A variance reduction technique is a statistical technique for improving the precision of a simulation out-put performance measure without using more simulation, or, alternatively achieve a desired precision with less simulation effort" (Kleijnen 1974).

It is know that the use of variance reduction techniques has potential benefits. However, the class of systems within which it is guaranteed to succeed and the particular technique that can achieve desirable magnitudes of variance reduction is ongoing research. In addition, applicability and success in the application of variance reduction techniques has been domain specific, without guidelines on their general use.

> "Variance reduction techniques cannot guarantee variance reduction in each simulation application, and even when it has been known to work, knowledge on the class of systems which it is provable to always work has remained rather limited" (Law and Kelton 2000).

The aim of this chapter is to answer the research question; which individual application of variance reduction techniques will succeed is achieving a reduction in variance for the different discrete event simulation scenarios under consideration. The scope of this chapter covers the use of variance reduction techniques as individual techniques on a set of scenarios from different application domains. The individual variance reduction techniques are:

   i.     Antithetic Variates
  ii.     Control Variates and
 iii.     Common Random Numbers.

In addition, the following three real world application domains are under consideration: (i) Manufacturing System (ii) Distribution System and (iii) Call Centre System. The rest of the book chapter is laid out as follows; the next section gives a background into the various concepts that underpin this study. This is followed by a case study section which describes the variance reduction techniques experimentation according to application domain. Further on is a discussion on the results from experimentation.

## 1.2  Reduction of Variance in Discrete Event Simulation

The development of simulation models requires a specific knowledge that is usually acquired over time and through experience. Since most simulation output results are essentially random variables, it may be difficult to determine whether an observation is as a result of system interrelationships or the randomness inherent in simulation models. Furthermore, simulation as a process can consume a lot of time, despite advances in computer technology. An example of a time consuming task is one which is statistically based i.e. output data analysis. However, it is known that advances in computer simulation have allowed the modeling of more complicated systems. Moreover, even when simpler systems are simulated, it can be difficult to judge the precision of simulation results. In general, output analysis is the examination of data generated by simulation experimentation, and its purpose is to predict the performance of a system or to compare the performance of two or more alternative system design (Law 2007).

However, simulation models differ from one another insofar as they have different values or types of system parameters, input variables, and behavioral relationships. These varying parameters, variables, and relationships are called "factors" and the output performance measure is called "response" in statistical design terminology (April et al. 2003). The decision as to which parameters are selected as fixed aspects of the simulation model and which are selected as experimental factors depends on the goals of the study rather than on the inherent form of the model. Also, during simulation studies there are usually a wide range of different responses or performance measure, which can be of interest. As a result, output performance measures for the three different simulation models considered within this study have been carefully selected after considering literature which reports on the most common performance metric for judging the performance of each simulation model (i.e. Manufacturing simulation, Call Centre simulation, and Cross-docking simulation). In addition, selection of output performance measures have been carried out in order to achieve a research goal of reducing simulation output variance through manual experimentation (Adewunmi 2010).

For simulation models, where the performance of such models is measured by its precision, i.e. mean, standard deviation, confidence interval and half width, for the selected output performance measure, it is sometimes difficult to achieve a target precision at an acceptable computational cost because of variance. This variance is usually that which is associated with the performance measure under consideration. For example, (Adewunmi et al. 2008), investigated the use of the Sequential Sampling Method (Law and Kelton 2000) to achieve a target variance reduction for a selected simulation output performance measure. Results from experimentation indicate that this technique for reducing variance requires a huge number of simulation runs to achieve any success for this particular simulation model. In a wider context, the variance associated with a simulation or its output performance measure may be due to the inherent randomness of the complex system under study. This variance can make it difficult to get precise estimates on the actual performance of the system. Consequently, there is a need to reduce the variance associated with the simulation output value, using the same or less simulation runs, in order to achieve a desired precision (Wilson 1984). The scope of this investigation covers the use of individual variance reduction techniques on different simulation models. This will be carried out under the assumption that all the simulation models for this study are not identical. The main difference between these models is the assumed level of inherent randomness. Where such randomness has been introduced by the following:

a. The use of probability distributions for modeling entity attributes such as inter arrival rate and machine failure. Conversely, within other models, some entity attributes have been modeled using schedules. The assumption is; the use of schedules does not generate as much randomness as with the use of probability distribution.
b. In addition, to the structural configuration of the simulation models under consideration i.e. the use of manual operatives, automated dispensing machines or a combination of both manual operatives and automated dispensing machines.

As a result, the manufacturing simulation model is characterized by an inter arrival rate and processing time which are modeled using probability distribution, the call centre simulation model's inter arrival rate and processing time are based on fixed schedules. The cross-docking simulation model is also characterized by the use of probability distribution to model the inter arrival rate and processing time of entities. The theoretical assumption is that by setting up these simulation models in this manner, there will be a variation in the level of model randomness. This should demonstrate the efficiency of the selected variance reduction techniques in achieving a reduction of variance for different simulation models, which are characterized by varying levels of randomness. In addition, as this is not a full scale simulation study, but a means of collecting output data for the variance reduction experiments, this investigation will not be following all the steps in a typical simulation study (Law 2007).

### 1.2.1 *Variance Reduction Techniques*

Within this section, the discussion has been restricted to a selected subset of variance reduction techniques which have proven to be the most practical in use within the discrete event simulation domain (Lavenberg and Welch 1978), (Cheng 1986). Furthermore, these techniques have been chosen because of the manner each one performs variance reduction i.e. through random number manipulation or the use of prior knowledge. The three selected variance reduction techniques fall into two broad categories; the first class manipulates random numbers for each replication of the simulation experiment, thereby inducing either a positive or a negative correlation between the mean responses across replications. Two methods of this category of variance reduction techniques are presented. The first method, Common Random Numbers, only applies when comparing two or more systems. The second method, using Antithetic Variates, applies when estimating the response of a variable of interest (Cole et al. 2001).

The second class of variance reduction techniques incorporates a modeler's prior knowledge of the system when estimating the mean response, which can result in a possible reduction in variance. By incorporating prior knowledge about a system into the estimation of the mean, the modeler's aim is to improve the reliability of the estimate. For this technique, it is assumed that there is some prior statistical knowledge of the system. A method that falls into this category is Control Variates (Nelson and Staum 2006). The following literature with extensive bibliographies is recommended to readers interested in going further into the subject i.e. (Nelson 1987), (Kleijnen 1988) and (Law 2007). In next section is a discussion on the three variance reduction techniques that appear to have the most promise of successful application to discrete event simulation modeling is presented.

### 1.2.1.1 Common Random Numbers (CRN)

Usually the use of CRN only applies when comparing two or more alternative scenarios of a single systems, it is probably the most commonly used variance reduction technique. Its popularity originates from its simplicity of implementation and general intuitive appeal. The technique of CRN is based on the premise that when two or more alternative systems are compared, it should be done under similar conditions (Bratley et al. 1986). The objective is to attribute any observed differences in performance measures to differences in the alternative systems, not to random fluctuations in the underlying experimental conditions. Statistical analysis based on common random numbers is founded on this single premise. Although a correlation is being introducing between paired responses, the difference, across pairs of replications is independent. This independence is achieved by employing a different starting seed for each of the pairs of replications. Unfortunately, there is no way to evaluate the increase or decrease in variance resulting from the use of CRN, other than to repeat the simulation runs without the use of the technique (Law and Kelton 2000).

There are specific instances where the use of CRN has been guaranteed. Gal et.al. present some theoretical and practical aspects of this technique, and discuss its efficiency as applied to production planning and inventory problems (Gal et al. 1984). In addition, Glasserman and Yao state that

> "common random numbers is known to be effective for many kinds of models, but its use is considered optimal for only a limited number of model classes".

They conclude that the application of CRN on discrete event simulation models is guaranteed to yield a variance reduction (Glasserman and Yao 1992). To demonstrate the concept of CRN, let $X_a$ denote the response for alternative $A$ and $X_b$ denote the response for alternative $B$, while considering a single system. Let $D$, denote the difference between the two alternatives, i.e. $D = X_a - X_b$. The following equation gives the random variable $D's$ variance.

$$Var(D) = Var(X_a X_b) + Var(X_a) - 2Cov(X_a, X_b) \qquad (1.1)$$

### 1.2.1.2 Antithetic Variates (AV)

In comparison to CRN, the AV technique reduces variance by artificially inducing a correlation between replications of the simulation model. Unlike CRN, the AV technique applies when seeking to improve the performance of a single system's performance. This approach to variance reduction makes $n$ independent pairs of correlated replications, where the paired replications are for the same system. The idea is to create each pair of replications such that a less than expected observation in the first replication is offset by a greater than expected observation in the second, and vice versa (Andreasson 1972), (Fishman and Huang 1983). Assuming that this value is closer to the expected response than the value that would result from the same number of completed independent replications, the average of the two observations is taken and the result used to derive the confidence interval.

A similar feature that AV shares with CRN is it can also be difficult to ascertain that it will work, and its feasibility and efficacy are perhaps even more model dependent than CRN. Another similarity it shares with CRN is the need for a pilot study to assess its usefulness in reducing variance for each specific simulation model (Cheng 1981). In some situations, the use of AV has been known to yield variance reduction, and as mentioned earlier it can be model specific. In his paper, Mitchell considers the use of AV to reduce the variance of estimates obtained in the simulation of a queuing system. The results reported in this paper, show that a reduction in variance of estimates was achieved (Mitchell 1973). The idea of AV is more formally presented. Let random variable $X$, denote the response from the first replication and $X'$ denote the replication from the second replication, within a pair. The random variable $Y$ denotes the average of these two variables, i.e. $Y = (X + X')/2$. The expected value of $Y$ and the variance of $Y$ are given as follows:

$$E(Y) = \frac{[E(X) + E(X')]}{2} = E(X) = E(X') \qquad (1.2)$$

and

$$Var(Y) = \frac{[Var(X) + Var(X') + 2Cov(X, X')]}{4} \qquad (1.3)$$

### 1.2.1.3 Control Variates (CV)

This technique is based on the use of secondary variables, called CV. This technique involves incorporating prior knowledge about a specific output performance parameter within a simulation model. It does not however require advance knowledge about a parameters theoretical relationship within the model as would other variance reduction techniques such as Indirect Estimation (IE). As compared with CRN and AV, CV attempts to exploit the advantage of the correlation between certain input and output variables to obtain a variance reduction. Of course depending on the specific type of CV that is being applied, the required correlation may arise naturally during the course of a simulation experiment, or might arise by using CRN in an auxiliary simulation experiment (Law 2007).

In order to apply the CV technique, it has to be assumed that a theoretical relationship exists between the control variate $X$, and the variable of interest $Y$. This approach does not require that a modeler knows the exact mathematical relationship between the control variates and the variable of interest; all the knowledge needed is to only know that the values are related. This relationship can be estimated by using the data recorded for instance from a pilot simulation study. Information from the estimated relationship is used to adjust the observed values of $Y$ (Sadowski et al. 1995). Let $X$ be the random variable that is said to partially control the random variable $Y$, and hence, it is called a control variate for $Y$. Usually it is assumed that there is a linear relationship between the variable of interest and the control variate. The observed values of the variable of interest $Y$ can then be corrected, by using the observed values of the control variates $X$, as follows:

$$Y_i(n) = Y(n) - a(X(n) - E(X)(n))$$  (1.4)

And

$$a = \frac{Cov(Y(n), X(n))}{Var(X)}$$  (1.5)

Where $a$ is the amount by which an upward or downward adjustment of the variable of interest $Y$ is carried out, $E(X)$ is the mean of $X$, and $n$ is the number of replications.

There are, however, some classes of discrete event simulation models for which the application of control variates has proven to be successful. In a recent article on the use of variance reduction techniques for manufacturing simulation by Eraslan and Dengiz, CV and Stratified Sampling were applied for the purpose of improving selected performance measures, results from this paper suggest that CV yields the lowest variance for selected performance measures (Eraslan and Dengiz 2009). The main advantage of using CV as a technique for variance reduction is that they are relatively easy to use. More importantly, CV can essentially be generated anywhere within the simulation run, so they add basically nothing to the simulation's cost; thus they will prove worthwhile even if they do not reduce the variance greatly (Kelton et al. 2007).

## 1.3 Case Studies

This section proceeds to present 3 case studies:

- The application of individual variance reduction techniques in a manufacturing system,
- The application of individual variance reduction techniques in a call centre system,
- The application of individual variance reduction techniques in a cross-docking distribution centre.

### *1.3.1 Manufacturing System*

#### 1.3.1.1 Description of a Manufacturing System / Simulation Model

Typically, the simulation of manufacturing systems is performed using a commercial software, rather than through a purpose built application. The manufacturing simulation model has been developed using the Arena$^{TM}$ simulation software. It is common that one of the activities during a simulation study is the statistical analysis of output performance measures. Since random samples from input probability distributions are used to model events in a manufacturing simulation model through time, basic simulation output data (e.g., average times in system of parts) or an estimated performance measure computed from them (e.g., average time in system from the entire simulation run) are also characterized by randomness (Buzacott and Yao 1986). Another source of manufacturing simulation model randomness which deserves a mention is unscheduled random downtime and machine failure which is also modeled using probability distributions. It is known that inherent model randomness can distort a true and fair view of the simulation model output results. Consequently, it is important to model system randomness correctly and also design and analyze simulation experiments in a proper manner (Law 2007).

There are a number of ways of modeling random unscheduled downtimes, interested readers are directed to Chapter 13, section 3, Discrete Event System Simulation, Banks et.al. (Banks et al 2000). The purpose of using variance reduction techniques is to deal with the inherent randomness in the manufacturing simulation model. This is through the reduction of variance associated with any selected measure of model performance. This reduction will be gained using the same number of replications that was used to achieve the initial simulation results. Improved simulation output results obtained from the application of variance reduction techniques has been known to increase the credibility of the simulation model.

An investigation into the application of variance reduction techniques on a small manufacturing simulation model is herein presented. The simulation model under consideration has been adapted from chapter 7, Simulation with Arena, Kelton et.al. (Kelton et al 2007), purely for research purposes. Experimentation is based on the assumption that the output performance measures are of a terminating, multi scenario, single system discrete event simulation model. The simple manufacturing system consists of parts arrival, four manufacturing cells, and parts departure. The system produces three part types, each routed through a different process plan in the system. This means that the parts do not visit individual Cells randomly, but through a predefined routing sequence. Parts enter the manufacturing system from the left hand side, and move only in a clockwise direction, through the system. There are four manufacturing cells; Cells 1, 2, and 4 each have a single machine, however, Cell 3 has two machines. The two machines at Cell 3 are not identical in performance capability; one of these machines is newer than the other and can perform 20% more efficiently than the other. Machine failure in Cells 1, 2, 3, and 4 in the manufacturing simulation model was represented using an exponential distribution with mean times in hours. Exponential distribution is a popular choice when modeling such activities in the absence of real data. A layout of the small manufacturing system under consideration is displayed in figure 1.1.

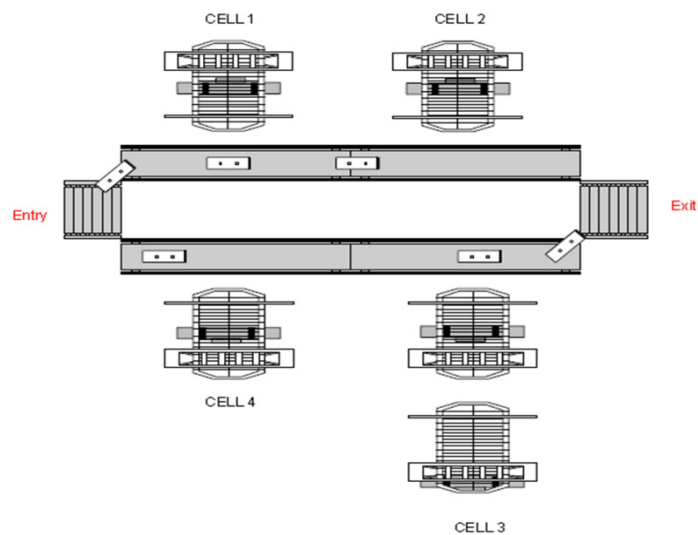

**Fig. 1.1** Small Manufacturing System Layout adapted from (Kelton et al. 2007) Chapter 7.

Herein is a description of the simulation model under consideration. All process times are triangularly distributed, while the inter arrival times between successive part arrivals are exponentially distributed. These are the probability distributions which were already implemented in the simulation model, and there was no reason not to continue using them. The Arena ™ simulation model incorporates an animation feature that captures the flow of parts to and fro the cells, until they are finally disposed or exist out of the system. The inter arrival times between successive parts arrival are exponentially distributed with a mean of 13 minutes, while the first part arrives at time 0.

Here is a brief description of the Arena^TM control logic which underlines the animation feature. Parts arrival are generated in the create parts module. The next step is the association of a routing sequence to arriving parts. This sequence will determine the servicing route of the parts to the various machine cells. Once a part arrives at a manufacturing cell (at a station), the arriving part will queue for a machine, and is then processed by a machine. This sequence is repeated at each of the manufacturing cells the part has to be processed. The process module for Cell 3 is slightly different from the other three Cells. This is to accommodate the two different machines, a new machine and an old machine, which process parts at different rates. Figure 1.2 shows the animation equivalent and control logic of the small manufacturing system simulation model.

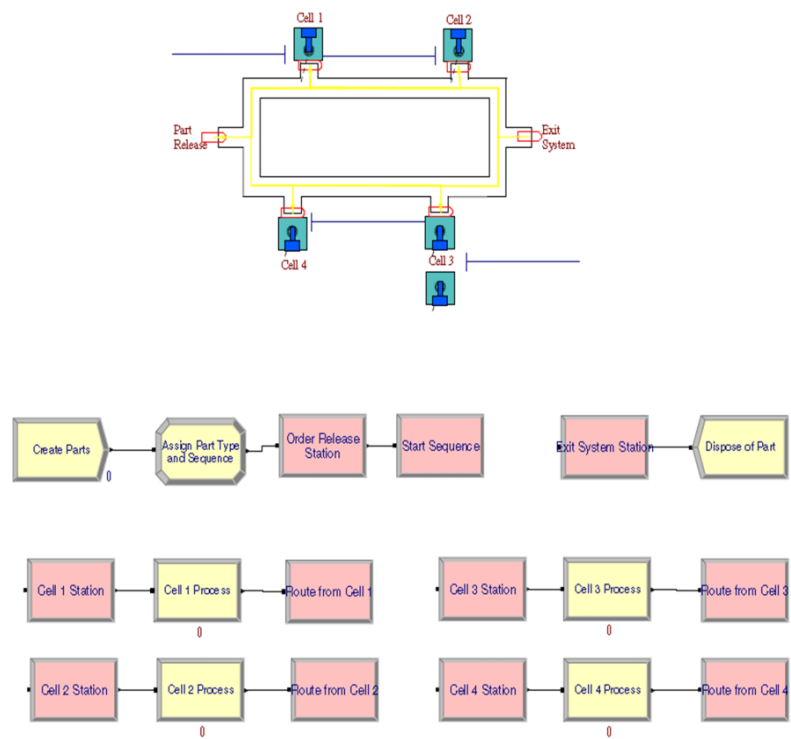

**Fig. 1.2** Manufacturing system simulation animation and control logic adapted from (Kelton et al. 2007) Chapter 7

### 1.3.1.2 Variance Reduction Experiments

This section of the chapter is divided into two parts; the first describes the design of the variance reduction experiments and the second details the results of the application of individual variance reduction techniques.

## 1.3.1.2.1  Experimental Design

In designing the variance reduction experiment, data on time persistent performance measures was utilized for experimentation as opposed to both time and cost data. This is due mainly to the availability of time based data as opposed to cost based data during the performance of the case study. Although both types of data would have given a greater insight into the performance of the variance reduction techniques, using different classes of time based data should be sufficient for this level of experimentation. Here is a list of the three performance measures utilized:

- Entity Total Average Time (Base): This is the average of the total time each entity will travel over the total length of the conveyor through the manufacturing system.
- Resource Utilization (Base): This variable records the instantaneous utilization of a resource during a specific period.
- Average Total WIP (Base): This metric records the average quantity of total work in process for each entity type.

The experimental conditions are as follows:

- Number of Replications: 10
- Warm up Period: 0
- Replication Length: 30 Days
- Terminating Condition: None

The performance measures have been labeled (Base), to highlight their distinction from those that have had variance reduction techniques applied and those that have not. As this is a pilot study where the goal is to establish the effectiveness of the variance reduction techniques under consideration, in this instance 10 simulation replications is deemed sufficient for collecting enough data for this purpose. An extensive bibliography on an appropriate number of replications for simulation experimentation and such like issues can be found in Robinson (Robinson 1994) and Hoad et.al (Hoad et al. 2009).In addition, for a full discussion on design issues such as warm up, replication length and simulation model termination condition for this study, readers are encouraged to see (Adewunmi 2010).

In addition, performance measures have been classed according to variance reduction techniques, i.e. Average Total WIP (Base), Average Total WIP (CRN), and Average Total WIP (AV). This means for each performance measure, the appropriate variance reduction that has been applied to it is stated, i.e. CRN and that which has not been treated to a variance reduction technique is labeled (Base). Under consideration is a two scenario, single manufacturing discrete event simulation model. The scenario which has performance measures labeled (Base) is characterized by random number seeds dedicated to sources of simulation model randomness as selected by the simulation software Arena $^{TM}$. The other scenario which has performance measures labeled common random number (CRN) has its identified sources of randomness, allocated dedicated random seeds by the user. So these two scenarios have unsynchronized and synchronized use of random numbers respectively (Law and Kelton 2000).

At this stage of experimental design, an additional performance measure Entity Wait Time is being introduced. This performance measure will be used for the CV experiment, with a view to applying it to adjusting upward or downwards the performance measure Entity Total Average Time (Base). Initial simulation results show a linear relationship between both variables, which will be exploited for variance reduction.

Here is the hypothesis that aim's to answer the research question:

- There is no difference in the standard deviations of the performance measure.

The hypothesis that tests the true standard deviation of the first scenario $\mu_1$ against the true standard deviation of the second scenario $\mu_2$,... scenario $\mu_k$ is:

$$H_0 : \mu_1 = \mu_2 = \ldots = \mu_k \tag{1.6}$$

Or

$$H_1 : \mu_i \neq \mu_k \ for \ at \ least \ one \ pair \ of \ (i, k) \tag{1.7}$$

Assuming we have samples of size $n_i$ from the $i - th$ population, $i = 1, 2, \ldots, k$, and the usual standard deviation estimates from each sample:

$$\mu_1, \mu_2 = \ldots = \mu_k \tag{1.8}$$

**Test Statistic: Bartlett's Test**

The Bartlett's Test (Snedecor and Cochran 1989) has been selected as a test for equality of variance between samples, as it is assumed that our data is normally distributed. Furthermore, this is one of the most common statistical techniques for this purpose. However, an alternative test like the Levene's test (Levene 1960) could have been used. In this instance, it will not be appropriate because Levene's test is less sensitive than the Bartlett test to departures from normality.

Significance Level: A value of $\alpha = 0.05$

Next is a summary of results from the application of individual variance reduction techniques on a manufacturing simulation model.

### 1.3.1.2.2 Results Summary

In this section, a summary of results on the performance of each variance reduction technique on each output performance measure is presented. In addition, a more in-depth description of results from the application of individual variance reduction techniques is presented in (Adewunmi 2010).

- At a 95% confidence interval (CI), homogeneity of variance was assessed by Bartlett's test. The P-value (0.000) is less than the significance level (0.05), therefore "reject the null hypothesis". The difference in variance between Average Total WIP (Base, CRN, AV, and CV) is "statistically significant". On the basis of the performance of the variance reduction techniques, CV technique achieved the largest reduction in variance for the simulation output performance measure, **Average Total WIP**.

- At a 95% confidence interval (CI), homogeneity of variance was assessed by Bartlett's test. The P-value (0.003) is less than the significance level (0.05), therefore "reject the null hypothesis". The difference in variance between Entity Total Average Time (Base, CRN, AV, and CV) is "statistically significant". On the basis of the performance of the variance reduction techniques, AV technique achieved the largest reduction in variance for the simulation output performance measure, **Entity Total Average Time**.
- At a 95% confidence interval (CI), homogeneity of variance was assessed by Bartlett's test. The P-value (0.006) is less than the significance level (0.05), therefore "reject the null hypothesis". The difference in variance between Resource Utilization (Base, CRN, AV, and CV) is "statistically significant". On the basis of the performance of the variance reduction techniques, CRN technique achieved the largest reduction in variance for the simulation output performance measure, **Resource Utilization**.

### 1.3.2 Call Centre System

#### 1.3.2.1 Description of a Call Centre System / Simulation Model

With the progression towards skill based routing of inbound customer calls due to advances in technology, Erlanger calculations for call centre performance analysis has become outdated since it assumes that agents have a single skill and there is no call priority (Doomun and Jungum 2008). On the other hand, the application of simulation ensures the modeling of human agent skills and abilities, best staffing decisions and provides an analyst with a virtual call centre that can be continually refined to answer questions about operational issues and even long term strategic decisions (L'Ecuyer and Buist 2006).

A close examination of a typical call centre reveals a complex interaction between several "resources" and "entities". Entities can take the form of customers calling into the call centre and resources are the human agents that receive calls and provide some service. These incoming calls, usually classified by call types, then find their way through the call centre according to a routing plan designed to handle specific incoming call type. While passing through the call centre, incoming calls occupy trunk lines, wait in one or several queues, abandon queues, and are redirected through interactive voice response systems until they reach their destination, the human agent. Otherwise, calls are passed from the interactive voice response system to an automatic call distributor (Doomun and Jungum 2008).

An automatic call distributor is a specialized switch designed to route each call to an individual human agent; if no qualified agent is available, then the call is placed in a queue. See figure 1.3 for an illustration of the sequence of activities in typical call centre, which has just been described in this section. Since each human agent possesses a unique skill in handling incoming calls, it is the customers' request that will determine whether the agent handles the call or transfers it to another agent. Once the call is handled, it then leaves the call centre system. During all of these call handling transactions, one critical resource being consumed is time. For example time spent handling a call and the time a call spends in the system. These are important metrics to consider during the evaluation of the performance of a call centre.

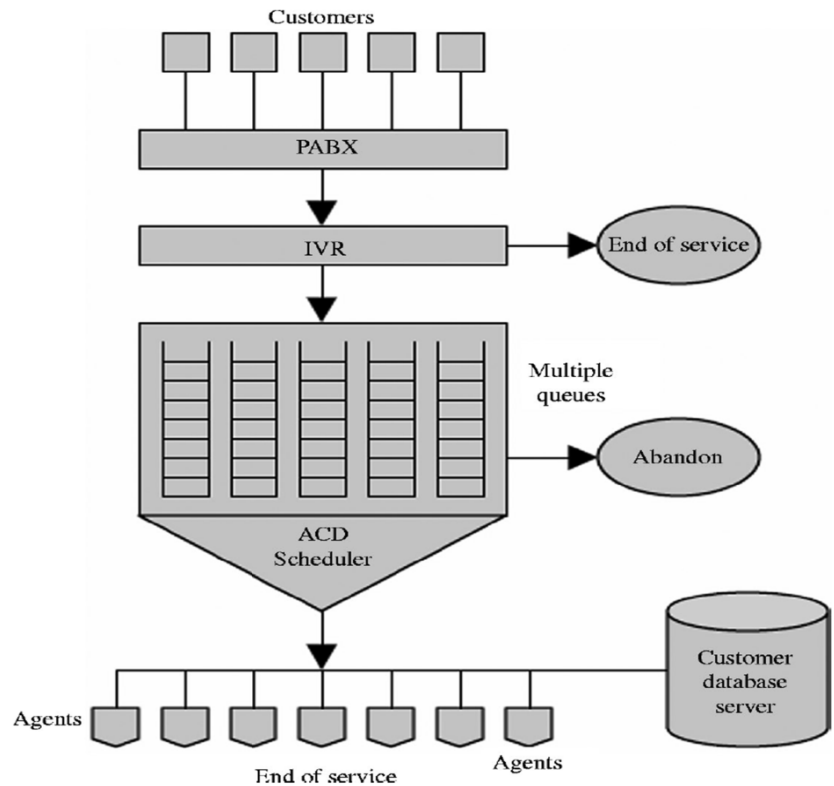

**Fig. 1.3** A Simple Call Centre adapted from (Doomun and Jungum 2008).

Herein is a description of the simulation model under consideration. The simple call centre system under consideration has been adapted from the Chapter 5, Simulation with Arena, (Kelton et al 2007). This call centre system, although theoretical in nature, contains the essential working components of a typical real life call centre, i.e. technical support, sales and customer order status checking. Arrival of incoming calls is generated using an arrival schedule. The purpose for using an arrival schedule instead of modeling this event using a probability distribution and a mean in minutes is to cause the system to stop creating new arrivals at a designated time into the simulation experiment. An answered caller has three options: transfer to technical support, sales information, or order status inquiry.

The estimated time for this activity is uniformly distributed; all times are in minutes.

In simulation terms, the "entities" for this simple call centre model are product type 1, 2 and 3. The available "resources" are the 26 trunk lines which are of a fixed capacity, and the sales and technical support staff. The skill of the sales and technical staff is modeled using schedules which show the duration during which for a fixed period, a resource is available, its capacity and skill level. The simulation model records the number of customer calls that are not able to get a trunk line and are thus rejected from entering the system similar to balking in queuing system. However, it does not consider "reneging", where customers who get a trunk line initially, later hang up the phone before being served. Figure 1.4, shows an Arena <sup>TM</sup> simulation animation of the simple call centre simulation model.

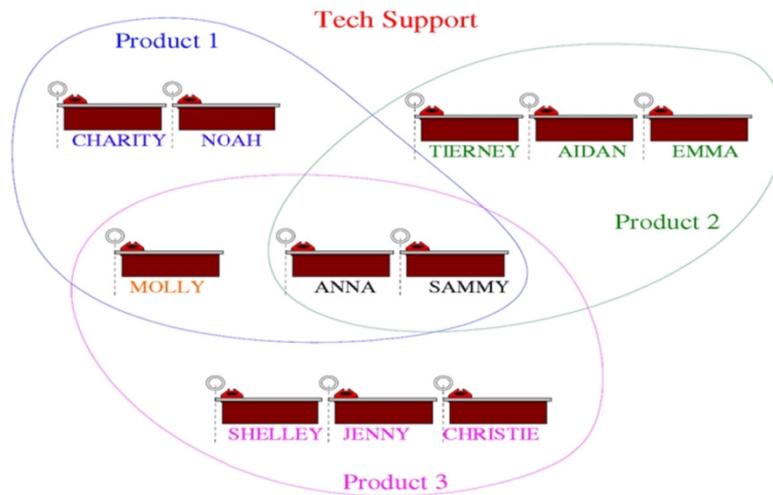

**Fig. 1.4** Call Centre Simulation Animation adapted from (Kelton et al. 2007) Chapter 5

### 1.3.2.2  Variance Reduction Experiments

This section of the chapter is divided into two parts; the first describes the design of the variance reduction experiments and the second details the results of the application of individual variance reduction techniques.

**Experimental Design**
For the design of the call centre variance reduction experiments, the three output performance measures which have been chosen are both time and cost persistent in nature. Here is a list of these performance measures:

- Total Average Call Time (Base): This output performance measure records the total average time an incoming call spends in the call centre simulation system.

- Total Resource Utilization (Base): This metric records the total scheduled usage of human resources in the operation of the call centre over a specified period in time.
- Total Resource Cost (Base): This is the total cost incurred for using a resource i.e. a human agent.

The experimental conditions are as follows:

- Number of Replications: 10
- Warm up Period: 0
- Replication Length: 660 minutes (27.5 days)
- Terminating Condition: At the end of 660 minutes and no queuing incoming

The call centre simulation model is based on the assumption that there are no entities at the start of each day of operation and the system will have emptied itself of entities at the end of the daily cycle. For the purpose of variance reduction experimentation, it is a terminating simulation model, although a call centre is naturally a non terminating system. No period of warm up has been added to the experimental set up. This is because experimentation is purely on the basis of a pilot run and the main simulation experiment, when it is performed, will handle issues like initial bias and its effect on the performance of variance reduction techniques. The performance measures have been labeled (Base), to highlight their distinction between those that have had variance reduction techniques applied and those that have not. These experiments assume that the sampled data is normally distributed.

In addition, the performance measures have been classed according to variance reduction techniques, i.e. Total Average Call Time (Base), Total Average Call Time (CRN), and Total Average Call Time (AV).Under consideration as in the previous manufacturing simulation study is a two scenario, single call centre simulation model. The scenario which has performance measures labeled (Base) is characterized by random number seeds dedicated to sources of simulation model randomness as selected by the simulation software Arena $^{TM}$. The other scenario which has performance measures labeled CRN has its identified sources of randomness, allocated dedicated random seeds by the user. So these two scenarios have unsynchronized and synchronized use of random numbers (Law and Kelton 2000).

The research question hypothesis remains the same as that in the manufacturing system; however an additional performance measure Total Entity Wait Time is introduced at this stage. This performance measure will be used for the CV experiment, with a view to adjusting the variance value of the performance measure Total Average Call Time (Base).

**Results Summary**

In this section, a summary of results on the performance of each variance reduction technique on each output performance measure is presented. In addition, a more in-depth description of results from the application of individual variance reduction techniques is presented in (Adewunmi 2010).

- At a 95% confidence interval (CI), homogeneity of variance was assessed by Bartlett's test. The P-value (0.000) is less than the significance level (0.05), therefore "reject the null hypothesis". The difference in variance between Total Aver-age Call Time (Base, CRN, AV, and CV) is "statistically significant". On the basis of the performance of the variance reduction techniques, CV technique achieved the largest reduction in variance for the simulation output performance measure, **Total Average Call Time**.
- At a 95% confidence interval (CI), homogeneity of variance was assessed by Bartlett's test. The P-value (0.995) is greater than the significance level (0.05), therefore "do not reject the null hypothesis". The difference in variance between Total Resource Utilization (Base, CRN, AV, and CV) is "statistically insignificant". On the basis of the performance of the variance reduction techniques, there was no reduction in variance for the simulation output performance measure, **Total Resource Utilisation**.
- At a 95% confidence interval (CI), homogeneity of variance was assessed by Bartlett's test. The P-value (0.002) is less than the significance level (0.05), therefore "reject the null hypothesis". The difference in variance between Total Re-course Cost (Base, CRN, AV, and CV) is "statistically significant". On the basis of the performance of the variance reduction techniques, AV technique achieved the largest reduction in variance for the simulation output performance measure, **Total Resource Cost**.

### *1.3.3 Cross-Docking System*

#### 1.3.3.1 Description of Cross-Docking System / Simulation Model

Many systems in areas such as manufacturing, warehousing and distribution can sometimes be too complex to model analytically; in particular, Just in Time (JIT) warehousing systems such as cross-docking can present such difficulty (Buzacott and Yao 1986). This is because cross-docking distribution systems operate processes which exhibit an inherent random behavior which can potentially affect its overall expected performance. A suitable technique for modeling and analyzing complex systems such as cross-docking systems is discrete event simulation (Magableh et al. 2005). Normally, such a facility would consist of a break up area where inbound freight is received and sorted as well as a build up area which handles the task of picking customer orders for onward dispatch via out bound dock doors. The usual activities of the cross-docking distribution centre begin with the receipt of customer orders, batched by outbound destinations, at specified periods during the day. As customer orders are being received, inbound freight arranged as pallet load is being delivered through inbound doors designated according to destination.

Customer orders batched by destination can differ in volume and variety; also they are released into the order picking system at the discretion of an operator in order to even out the work load on the order picking system. Once pallet load is sorted by a floor operative i.e. during the break up process, individual items in packs of six to twelve units can be placed in totes (A plastic container which is used for holding items on the conveyor belt). Normally, totes will begin their journey on a conveyor belt, for onward routing to the order picking area. Just before the order picking area is a set of roof high shelves where stock for replenishing the order picking area is kept. A conveyor belt runs through the order picking area and its route and speed are fixed. Figure 1.5, below provides a representation of the cross-docking distribution centre.

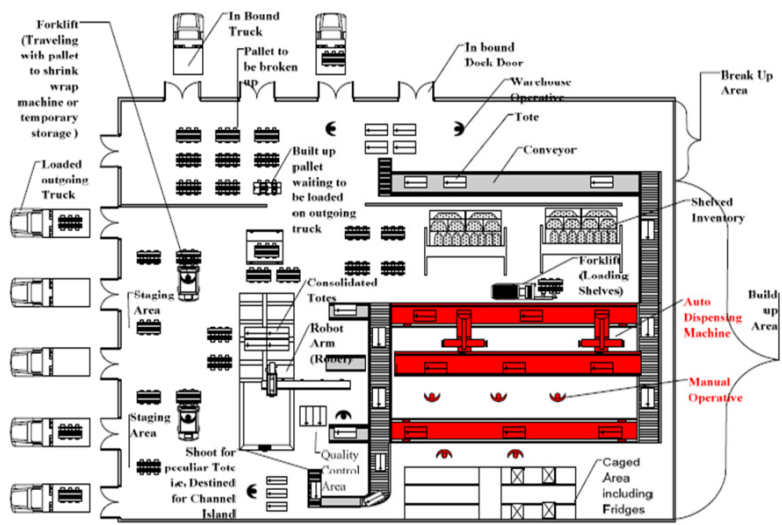

**Fig. 1.5** A Typical Cross-docking Distribution Centre (Adewunmi 2010).

Within the order picking area, there are two types of order picking methods; automated dispensing machines and manual order picking operatives. These order picking resources are usually available in shifts, constrained by capacity and scheduled into order picking jobs. There is also the possibility that manual order picking operators possess different skill levels and there is a potential for automated order picking machines to breakdown. In such a situation, it becomes important for the achievement of a smooth cross-docking operation, to pay particular attention to the order picking process within the cross-docking distribution system. The order picking process essentially needs to be fulfilled with minimal interruptions and with the least amount of resource cost (Lin and Lu 1999). Below figure 1.6 provides a representation of the order picking function with a cross-docking distribution centre.

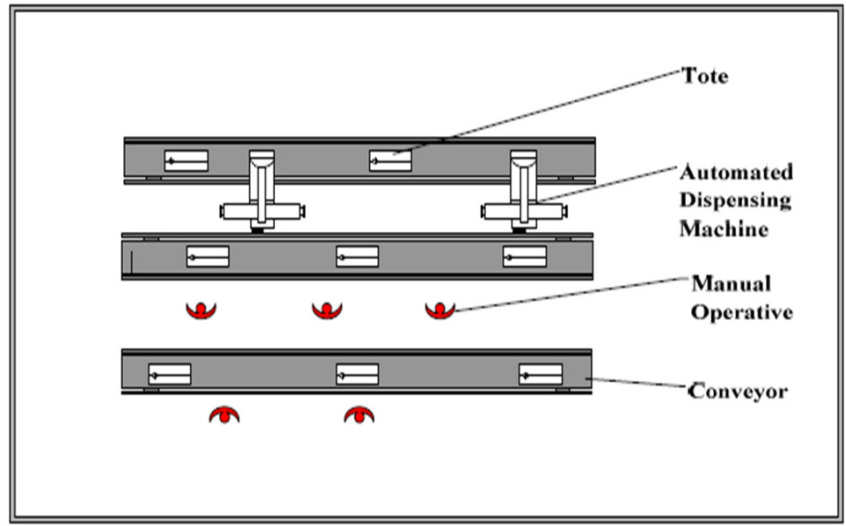

**Fig. 1.6** An Order Picking Process within a Cross-docking Distribution Centre (Adewunmi 2010)

A description of the order picking simulation model, which will be the scope of the cross-docking simulation study is presented. The scope of this particular study is restricted to the order picking function as a result of an initial investigation conducted at a physical cross-docking distribution centre. It was discovered that amongst the different activities performed in a distribution centre, the order picking function was judged as the most significant by management. The customer order (entity) inter arrival rate is modeled using an exponential probability distribution, and the manual as well as the automated order picking process are modeled using triangular probability distribution. Customer orders are released from the left hand side of the simulation model. At the top of the model are two automated dispensing machines and at the bottom of the simulation model are two sets of manual order picking operatives, with different levels of proficiency in picking customer orders. Figure 1.7, displays a simulation animation of the order picking process cross-docking distribution centre.

### 1.3.3.2  Variance Reduction Experiments

This section of the chapter is divided into two parts; the first describes the design of the variance reduction experiments and the second details the results of the application of individual variance reduction techniques.

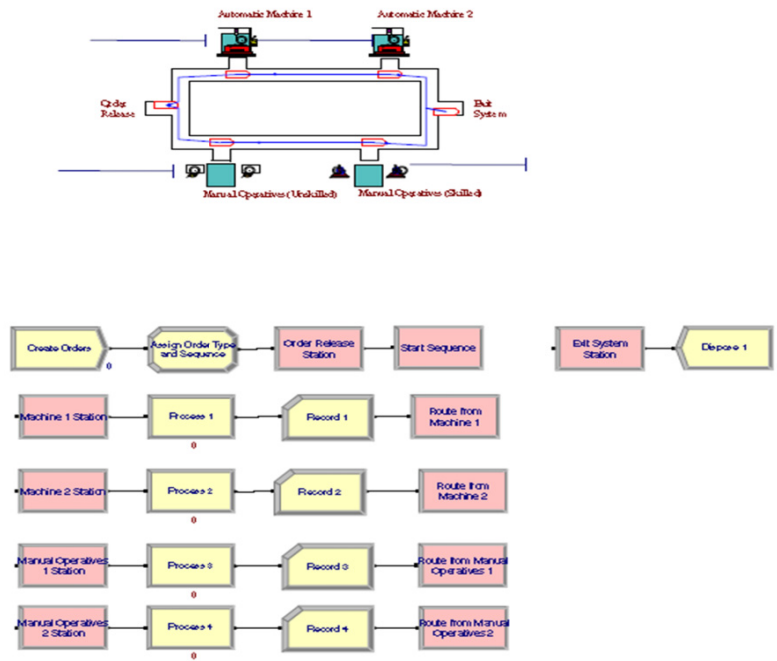

**Fig. 1.7** Simulation animation of a Cross-docking order picking process (Adewunmi 2010).

### Experimental Design

For the design of the cross-docking distribution system variance reduction experiments, the following performance measures were chosen:

- Total Entity Time (Base): This variable records the total time an entity spends in the simulation system.
- Total Resource Utilization (Base): The purpose of collecting data on resource utilization is to have statistics on the level of usage of the resources during a specified period.
- Total Resource Cost (Base): This is a cost based statistic that records the monetary amount expended on the use of re-sources for a specific period.

The experimental conditions are as follows:

- Number of Replications: 10
- Warm up Period: 0
- Replication Length: 30 Days
- Terminating Condition: None

The performance measures have been classed according to variance reduction technique, i.e. Total Resource Utilization (Base), Total Resource Utilization (CRN), and Total Resource Utilization (AV). Under consideration is a two scenario, single cross-docking discrete event simulation model. The scenario which has performance measures labeled (Base) is characterized by random number seeds dedicated to sources of simulation model randomness as selected by the simulation software Arena [TM]. The other scenario which has performance measures labeled CRN has its identified sources of randomness, allocated dedicated random seeds by the user. So these two scenarios have unsynchronized and synchronized use of random numbers (Law and Kelton 2000).

The research question hypothesis remains the same as that in the manufacturing system; however an additional performance measure Total Entity Wait Time is introduced at this stage. This performance measure will be used for the CV experiment, with a view to applying it to adjusting the performance measure Total Entity Time. For those interested, detailed results from the application of individual variance reduction techniques are presented in (Adewunmi 2010).

**Results Summary**

In this section, a summary of results on the performance of each variance reduction technique on each output performance measure is presented. In addition, a more in-depth description of results from the application of individual variance reduction techniques is presented in (Adewunmi 2010).

- At a 95% confidence interval (CI), homogeneity of variance was assessed by Bartlett's test. The P-value (0.000) is less than the significance level (0.05), therefore "reject the null hypothesis". The difference in variance between Total Entity Time (Base, CRN, AV, and CV) is "statistically significant". On the basis of the performance of the variance reduction techniques, CV technique achieved the largest reduction in variance for the simulation output performance measure, **Total Entity Time**.
- At a 95% confidence interval (CI), homogeneity of variance was assessed by Bartlett's test. The P-value (0.000) is less than the significance level (0.05), therefore "reject the null hypothesis". The difference in variance between Total Re-source Cost (Base, CRN, AV, and CV) is "statistically significant". On the basis of the performance of the variance reduction techniques, AV technique achieved the largest reduction in variance for the simulation output performance measure, **Total Resource Cost**.
- At a 95% confidence interval (CI), homogeneity of variance was assessed by Bartlett's test. The P-value (0.003) is less than the significance level (0.05), therefore "reject the null hypothesis". The difference in variance between Total Resource Utilization (Base, CRN, AV, and CV) is "statistically significant". On the basis of the performance of the variance reduction techniques, AV technique achieved the largest reduction in variance for the simulation output performance measure, **Total Resource Utilization**.

## 1.4 Discussion

The purpose of this study is to investigate the application of variance reduction techniques (CRN, AV and CV) on scenarios from three different application domains. In addition, to finding out which class of systems the variance reduction techniques will prove to most likely succeed. It also seeks to provide general guidance to beginners on the universal applicability of variance reduction techniques. A review of results from the variance reduction experiments indicate that the amount of variance reduction by the techniques applied can vary substantially from one output performance measure to the other, as well as one simulation model to the other. Among the individual techniques, CV stands out as the best technique. This is followed by AV and CRN. CV was the only technique that achieved a reduction in variance for at least one performance measure of interest, in all three application domains. This can be attributable to the fact that the strength of this technique is its ability to generate a reduction in variance by inducing a correlation between random variates. In addition, control variates have the added advantage of being able to be used on more than one variate, resulting in a greater potential for variance reduction. However, implementing AV and CRN required less time, and was less complex than CV for all three domain application domains. This maybe because with CV, where there is a need to establish some theoretical relationship between the control variate and the variable of interest.

The variance reduction experiments were designed with the manufacturing simulation model being characterized by an inter arrival rate and processing time which were modeled using probability distribution. The cross-docking simulation model was also characterized by the use of probability distribution to model the inter arrival rate and processing time of entities. Conversely, the call centre simulation model inter arrival rate and processing time were based on fixed schedules. The assumption is that by setting up these simulation models in this manner, there will be a variation in the level of model randomness i.e. the use of schedules does not generate as much model randomness as with the use of probability distribution. For example, results demonstrate that for the call centre simulation model, the performance measure "Total Resource Utilization" did not achieve a reduction in variance with the application of CRN, AV and CV, on this occasion. However, for this same model, the performance measures "Total Average Call Time" and "Total Resource Cost" did achieve a reduction in variance. This expected outcome demonstrates the relationship between the inherent simulation model's randomness and the efficiency of CRN, AV and CV, which has to be considered when applying variance reduction techniques in simulation models.

This study has shown that the Glasserman and Yao (Glasserman and Yao 1992) statement regarding the general applicability of CRN is true, for the scenarios and application domains under consideration. As a consequence, this makes CRN a more popular choice of technique in theory. However, results from this study demonstrate CRN to be useful but not the most effective technique for reducing variance. In addition CV under the experimental conditions reported within this study did outperform CRN. While it is not claimed that CV is more superior a technique as compared with CRN, in this instance, it has been demonstrated that CV achieved more instances of variance reduction as compared with CRN and AV. In addition, under current experimental conditions, a new specific class of systems, in particular the Cross-docking distribution system has been identified, for which the application of CV and AV is beneficial for variance reduction.

## 1.5 Conclusion

Usually during a simulation study, there are a variety of decisions to be made at the pre and post experimentation stages. Such decisions include input analysis, design of experiments and output analysis. Our interest is in output analysis with particular focus on the selection of variance reduction techniques as well as their applicability. The process of selection was investigated through the application of CRN, AV and CV in a variety of scenarios. In addition, this study seeks to establish which of the application domains considered, will the application of CRN, AV and CV be successful, where such success had not been previously reported. Amongst the individual variance reduction techniques (CRN, AV and CV), CV was found to be most effective for all the application domains considered within this study. Furthermore, AV and CV, individually, were effective in variance reduction for the cross-docking simulation model. Typically, a lot of consideration is given to number of replications, replication length, terminating condition, warm up period during the design of a typical simulation experiment. It would be logical to imagine that there will be a linear relationship between these factors and the performance of variance reduction techniques. However, the extent of this relationship is unknown unless a full simulation study is performed before the application of variance reduction techniques. The experimental conditions applied to this study were sufficient to demonstrate reduction. However, upcoming research will investigate the nature and effect of considering the application of variance reduction techniques during the design of experiments for full scale simulation study.

In future, research investigation will be focused on exploring the idea of combining different variance reduction techniques, with the hope that their individual beneficial effort will add up to a greater magnitude of variance reduction for the estimator of interest. These combinations could have a positive effect when several alternative configurations are being considered. To obtain more variance reduction, one may want to combine variance reduction techniques simultaneously in the same simulation experiment and use more complicated discrete event simulation models. The potential gain which may accrue from the combination of these techniques is also worth investigating because it will increase the already existing knowledge base on such a subject.

## Authors Biography, Contact

Dr Adrian Adewunmi was a Post Graduate Researcher in the Intelligent Modelling & Analysis (IMA) Research Group, School of Computer Science, University of Nottingham. A summary of his current interest is Modeling and Simulation, Artificial Intelligence and Data Analysis.´

Professor Uwe Aickelin is an EPSRC Advanced Research Fellow and Professor of Computer Science at The University of Nottingham. He is also the Director of Research in the School of Computer Science and leads one of its four research groups: Intelligent Modeling & Analysis (IMA). A summary of his current research interests is Modeling and Simulation, Artificial Intelligence and Data Analysis.

**Contact**

adrian.a.adewunmi@googlemail.com
uwe.aickelin@nottingham.ac.uk
Intelligent Modelling & Analysis Research Group (IMA)
School of Computer Science
The University of Nottingham
Jubilee Campus
Wollaton Road
Nottingham NG8 1BB
UK